\documentclass[aps,prd,twocolumn,showpacs,groupedaddress,floatfix]{revtex4}

\usepackage{graphicx}  
\usepackage{dcolumn}   
\usepackage{bm}        
\usepackage{amssymb}   
\usepackage{multirow}
\usepackage{xspace}

\def\lsim{\mathrel{\rlap{\lower4pt\hbox{\hskip1pt$\sim$}}
    \raise1pt\hbox{$<$}}}                

\begin{document}

\newcommand{\dzero}     {D0}
\newcommand{\ttbar}     {\mbox{$t\bar{t}$}}
\newcommand{\bbbar}     {\mbox{$b\bar{b}$}}
\newcommand{\ccbar}     {\mbox{$c\bar{c}$}}
\newcommand{\pythia}    {{\sc{pythia}}}
\newcommand{\alpgen}    {{\sc{alpgen}}}
\newcommand{\geant}     {{\sc geant3}}
\newcommand{\metcal}    {\mbox{$\not\!\!E_{Tcal}$}}
\newcommand{\met}       {\mbox{$\not\!\!E_T$}}
\newcommand{\pt}	{$p_T$}
\newcommand{\mtop}      {$m_t$}

\newcommand{\lumi}      {1~$\rm fb^{-1}$}
\newcommand{\result}    {7.5}
\newcommand{\erstat}    {^{+1.0}_{-1.0}}
\newcommand{\ersyspos}  {+0.7}
\newcommand{\ersysneg}  {-0.6}
\newcommand{\ersys}     {^{\ersyspos}_{\ersysneg}}
\newcommand{\erlumi}    {^{+0.6}_{-0.5}}
\newcommand{\fullresult} {\ensuremath{\result \erstat\,{\rm(stat)}\,\ersys\,{\rm(syst)}\,\erlumi\,{\rm(lumi)}}\xspace}

\newcommand{\resultll}    {7.5}
\newcommand{\erstatll}    {^{+1.2}_{-1.1}}
\newcommand{\ersysllpos}  {+0.7}
\newcommand{\ersysllneg}  {-0.6}
\newcommand{\ersysll}     {^{\ersyspos}_{\ersysneg}}
\newcommand{\erlumill}    {^{+0.7}_{-0.5}}
\newcommand{\fullresultll} {\ensuremath{\resultll \erstatll\,{\rm(stat)}\,\ersysll\,{\rm(syst)}\,\erlumill\,{\rm(lumi)}}\xspace}

\newcommand{\lumiee}      {1.07~$\rm fb^{-1}$}
\newcommand{\resultee}    {9.6}
\newcommand{\erstatee}    {^{+3.2}_{-2.7}}
\newcommand{\ersysposee}  {+1.0}
\newcommand{\ersysnegee}  {-0.9}
\newcommand{\ersysee}     {^{\ersysposee}_{\ersysnegee}}
\newcommand{\erlumiee}    {^{+0.8}_{-0.7}}
\newcommand{\observedevtsee} {17}
\newcommand{\totbkgee}    {$3.4\pm0.8$}
\newcommand{\fullee}      {\ensuremath{\resultee \erstatee\,{\rm(stat)}\,\ersysee\,{\rm(syst)}\,\erlumiee\,{\rm(lumi)}}\xspace}

\newcommand{\lumiemu}      {1.07~$\rm fb^{-1}$}
\newcommand{\resultemu}    {7.2}
\newcommand{\erstatemu}    {^{+1.4}_{-1.3}}
\newcommand{\ersysposemu}  {+0.8}
\newcommand{\ersysnegemu}  {-0.7}
\newcommand{\ersysemu}     {^{\ersysposemu}_{\ersysnegemu}}
\newcommand{\erlumiemu}    {0.6}
\newcommand{\observedevtsemu} {39}
\newcommand{\totbkgemu}    {$9.5\pm1.2$}
\newcommand{\fullemu}      {\ensuremath{\resultemu \erstatemu\,{\rm(stat)}\,\ersysemu\,{\rm(syst)}\,\pm\,\erlumiemu\,{\rm(lumi)}}\xspace}

\newcommand{\lumimumu}      {1.01~$\rm fb^{-1}$}
\newcommand{\resultmumu}    {5.1}
\newcommand{\erstatmumu}    {^{+3.4}_{-2.8}}
\newcommand{\ersysposmumu}  {+1.3}
\newcommand{\ersysnegmumu}  {-1.1}
\newcommand{\ersysmumu}     {^{\ersysposmumu}_{\ersysnegmumu}}
\newcommand{\erlumimumu}    {^{+0.7}_{-0.6}}
\newcommand{\observedevtsmumu} {12}
\newcommand{\totbkgmumu}    {$6.4\pm0.5$}
\newcommand{\fullmumu}      {\ensuremath{\resultmumu \erstatmumu\,{\rm(stat)}\,\ersysmumu\,{\rm(syst)}\,\erlumimumu\,{\rm(lumi)}}\xspace}

\newcommand{\mtc}	{171.5}
\newcommand{\ermtc}	{^{+9.9}_{-8.8}}
\newcommand{\mtm}       {173.3}
\newcommand{\ermtm}     {^{+9.8}_{-8.6}}

\hspace{5.2in} \mbox{FERMILAB-PUB-09-013-E}
\title{Measurement of the {\boldmath $t\bar{t}$} production cross section and top quark mass extraction 
  using dilepton events in {\boldmath $p\bar{p}$} collisions}
\vspace*{0.1cm}

%
\author{V.M.~Abazov$^{36}$}
\author{B.~Abbott$^{75}$}
\author{M.~Abolins$^{65}$}
\author{B.S.~Acharya$^{29}$}
\author{M.~Adams$^{51}$}
\author{T.~Adams$^{49}$}
\author{E.~Aguilo$^{6}$}
\author{M.~Ahsan$^{59}$}
\author{G.D.~Alexeev$^{36}$}
\author{G.~Alkhazov$^{40}$}
% ...

\affiliation{\vspace{0.1 in}(The D\O\ Collaboration)\vspace{0.1 in}}
% ...
\affiliation{$^{6}$University of Alberta, Edmonton, Alberta, Canada,
                Simon Fraser University, Burnaby, British Columbia, Canada,
                York University, Toronto, Ontario, Canada, and
                McGill University, Montreal, Quebec, Canada}
% ...
\affiliation{$^{29}$Tata Institute of Fundamental Research, Mumbai, India}
% ...
\affiliation{$^{36}$Joint Institute for Nuclear Research, Dubna, Russia}
% ...
\affiliation{$^{40}$Petersburg Nuclear Physics Institute,
                St. Petersburg, Russia}
% ...
\affiliation{$^{49}$Florida State University, Tallahassee, Florida 32306, USA}
% ...
\affiliation{$^{51}$University of Illinois at Chicago,
                Chicago, Illinois 60607, USA}
% ...
\affiliation{$^{59}$Kansas State University, Manhattan, Kansas 66506, USA}
% ...
\affiliation{$^{65}$Michigan State University,
                East Lansing, Michigan 48824, USA}
% ...
\affiliation{$^{75}$University of Oklahoma, Norman, Oklahoma 73019, USA}
% ...


\date{July 14, 2009}

\begin{abstract}
We present a measurement of the top quark pair production cross section 
in $p\bar{p}$ collisions at $\sqrt{s}=1.96$~TeV using 
approximately 1~fb$^{-1}$ collected with the D0 detector.  
We consider decay channels containing two high \pt\ charged leptons
where one lepton is identified as an electron or a muon while the other
lepton can be an electron, a muon or a hadronically decaying $\tau$ lepton.
For a mass of the top quark of 170~GeV, the measured cross section is \fullresult pb.
Using $\ell \tau$ events only, we measure: 
$\sigma_{t\bar{t}} \times B(\ttbar \to \ell \tau b\bar{b}) = 0.13
   ^{+0.09}_{-0.08}\text{(stat)}
   ^{+0.06} _{-0.06} (\mathrm{syst})
   ^{+0.02} _{-0.02} (\mathrm{lumi}) \mathrm{~~pb}$.
Comparing the measured cross section as a function of the mass of the top quark
with a partial next-to-next-to leading order Quantum Chromodynamics 
theoretical prediction, we extract a mass of the top quark of  
$\mtc \ermtc $~GeV, in agreement with direct measurements.
\end{abstract}


\maketitle 


The top quark, first observed at Fermilab in
1995~\cite{top_discovery_cdf, top_discovery_d0}, 
is the heaviest known
elementary particle. In many extensions of the standard model (SM) new
physics is predicted in connection with top quarks. 
In the SM, top quarks are predicted to decay into a $W$~boson and a
$b$~quark with a branching fraction of nearly 100\%~\cite{Yao:2006px}.
For approximately $10\%$ of all top-antitop quark (\ttbar) events, both
$W$ bosons decay leptonically and generate final states containing two leptons~\cite{Yao:2006px}.
In addition, these final states are characterized by the presence of two high energy jets 
resulting from hadronization of the two $b$~quarks and large
imbalance in transverse momentum ($\met$) due to several undetected neutrinos
from the $W$ boson decays.

New physics in the production or decay of the top quark may lead to significant deviations 
in the measured \ttbar\ cross section ($\sigma_{t\bar{t}}$) from the SM prediction.
Since new physics could have a different impact on
different final states, it is
important to measure $\sigma_{t\bar{t}}$ precisely  
in all possible decay channels.
Channels including a $\tau$ lepton in the final state are of particular interest, 
since the decay chain involves only third generation fermions.
Owing to the significant dependence of the \ttbar\ cross section on the mass of the top quark (\mtop), 
a precise cross section measurement allows the extraction of the mass of the top quark in a way
complementary to direct reconstruction methods and hence provides a valuable consistency check.
A measurement of the mass of the top quark is important
since together with that of the $W$ boson, it allows one to place indirect
constraints on the mass of the SM Higgs boson. 

In this Letter, we present a measurement of $\sigma_{t\bar{t}}$ using
approximatively 1~fb$^{-1}$ 
from Run II of the Fermilab Tevatron $p\bar{p}$ collider operated at $\sqrt{s}=1.96$~TeV,
and collected with the D0 detector. We consider dilepton final states with two
identified electrons or muons from the $W$ boson leptonic decays, i.e., $ee, e\mu$ and $\mu\mu$, and 
final states with a $\tau$ lepton that decays into hadrons+$\nu_{\tau}$ from the decay of one  $W$ boson and
an accompanying electron or muon from the other $W$ boson, i.e., $e\tau$ and $\mu\tau$.
Throughout the text, these final states will be referred to as $\ell \ell$ and $\ell \tau$ channels,
respectively.
Dilepton channels also have contributions from events where both $\tau$ leptons 
decay into electrons or muons.
Previous measurements of $\sigma_{t\bar{t}}$ in the dilepton channel were reported
in~\cite{Abachi:2007:prd,cdf:dilepton}. 
We update the \dzero\ measurement~\cite{Abachi:2007:prd} using more
integrated luminosity and include the $\ell \tau$ final states in the result.
We also present a measurement of $\sigma_{t\bar{t}} \times B(\ttbar \to \ell \tau b\bar{b})$.
In addition, we explore the dependence of $\sigma_{t\bar{t}}$ on the mass of the top quark,
and through a comparison
with higher order Quantum Chromodynamics (QCD) calculations computed in the
pole mass scheme, we extract a value for the mass of the top quark.


The \dzero\ detector has a central tracking system, consisting of a
silicon microstrip tracker (SMT) and a central fiber tracker,
both located within a 2~T superconducting solenoidal
magnet~\cite{Abazov:2005pn}.
A liquid argon and uranium calorimeter has a
central section covering pseudorapidities $|\eta|$
up to $\approx 1.1$~\cite{pseudorap}, and two end calorimeters (EC) that extend coverage
to $|\eta|\approx 4.2$, with all three calorimeters housed in separate
cryostats~\cite{Abachi:1993em}. An outer muon system, covering $|\eta|<2$,
consists of a layer of tracking detectors and scintillation trigger
counters in front of 1.8~T iron toroids, followed by two similar layers
after the toroids~\cite{munim}. The luminosity is measured using plastic scintillator
arrays placed in front of the EC cryostats. The trigger and data
acquisition systems are designed to accommodate the high luminosities
of Run~II. The dilepton triggers used in the $\ell \ell$ channels are described in Ref.~\cite{Abachi:2007:prd}.  
The $\ell \tau$ channel uses triggers requiring one lepton and one jet.
The trigger efficiency for signal events passing the selection and acceptance cuts 
varies from 78\% to 98\% depending on the channel. 


Electrons are identified as clusters of energy deposits in calorimeter 
cells satisfying the following requirements: (i) the fraction of energy deposited  in the electromagnetic
section of the calorimeter is at least 90\% of the total cluster energy,
(ii) the energy is concentrated in a narrow cone, and isolated from
other energy deposits, (iii) the shape of the shower is compatible
with that of an electron, and (iv) a track extrapolated from the tracking
system points to the cluster. To further reduce backgrounds (see below for background description)
we use a likelihood discriminant that selects prompt isolated electrons, based on tracking 
and calorimetric information. Both central~($|\eta|<1.1$) and 
forward~($1.5<|\eta|<2.5$) electron candidates are accepted.  

Muon trajectories are reconstructed using hits in three
layers of the outer muon system along with matching tracks in the inner tracker.
The energy deposited within an annulus
$0.1 < \sqrt{(\Delta \eta)^2 + (\Delta\phi)^2} < 0.4$ around the muon direction (where 
$\phi$ is the azimuthal angle) must 
be less than 15\% of the muon \pt, for all channels except $\mu\mu$, while for 
the $\mu\mu$ final state, the selected muons must not lie within the
cone of any reconstructed jet. 
To reduce background further, the
sum of the track momenta in a cone around the
muon track has to be smaller than 15\% of the muon \pt.
Moreover, the fraction of prompt muons is increased
by requiring that the distance of closest approach of the muon track 
to the primary vertex is small.

A hadronically decaying $\tau$ lepton is characterized by a narrow
jet of low track multiplicity. The $\tau$ lepton reconstruction is seeded either by
a calorimeter energy cluster using the D0 Run II cone algorithm~\cite{jetcone} 
with radius ${\cal R}~=~0.3$ or by a track. Three types of $\tau$ decays are defined
as (i) $\tau$-type 1 ($\pi^{\pm}$-like), consisting of a single track, 
with energy deposition in the hadronic calorimeter, (ii) $\tau$-type 2
($\rho^{\pm}$-like), a single track, with an energy deposit in both the hadronic
and the electromagnetic calorimeters and (iii) $\tau$-type 3,
having two or three tracks, forming an invariant mass $<1.1$~or $<1.7$~GeV, respectively.\
The total sum of the particle charges for $\tau$-type 3 is required to be $\pm 1$ or $\pm 2$.
A set of neural networks ({\it NN}$_{\tau}$), one for each $\tau$-type, has been developed based on 
discriminating variables discussed in Ref.~\cite{nntau}. These variables exploit differences
between hadronically decaying $\tau$ leptons and jets resulting from the fragmentation of
quarks and gluons, in particular the longitudinal and 
transverse shower shapes as well as isolation in the calorimeter 
and in the tracker. This technique has been used to perform a measurement of 
$\sigma(p\bar{p}\to Z+X)\times BR(Z\to\tau^+\tau^-)$~\cite{ztautauplb}.

Jets are reconstructed using a fixed cone algorithm with radius 
${\cal R}~=~0.5$~\cite{jetcone}.
A jet energy scale calibration obtained from transverse momentum balance in $\gamma$+jet events 
is applied to all jets. \met\ is defined as equal in magnitude 
and opposite in direction to the vector sum of all significant
transverse energies in calorimeter cells.
It is further corrected by the transverse momentum of all reconstructed muons,
as well as by the energy calibration corrections applied to the transverse momenta
of electrons, $\tau$ leptons and jets.
A more detailed description of object reconstruction can be found
in Ref.~\cite{Abachi:2007:prd}.

Jets from $b$ quarks are identified using a neural network $b$ jet tagging
algorithm~\cite{bid}. It combines several variables that characterize the presence and 
properties of the secondary
vertices and the tracks of high impact parameter within jets.
We obtain a 54\% average tagging efficiency in data for $b$ jets containing at least two tracks 
with SMT hits~\cite{bid},
which corresponds to a 1\% mistagging of jets from light quark flavors ($u,d$ or $s$ quarks) as $b$ jets.
The identification of $b$ jets is only used in the $\ell \tau$ channel.


In the $\ell \ell$ channels, the main source of background is the production of electroweak
bosons that decay to charged leptons.
It arises from $Z/\gamma^{\ast} \rightarrow \ell^+\ell^-$ and
$Z/\gamma^{\ast} \rightarrow \tau^+\tau^-$, followed by $\tau \rightarrow \ell^\pm \nu_\ell \nu_\tau$
with $\ell^\pm=e^\pm$~or~$\mu^\pm$, along with diboson production  
($WW$, $WZ$ and $ZZ$), when the boson decays lead to at least two charged leptons in the final state. 
In the $\ell \tau$ channel, the dominant background emerges from jets mimicking electrons and $\tau$ leptons, 
muons from semileptonic $b$ quark decay or pion or kaon decay,
and large misreconstructed \met, mainly in $W$+jets and multijet production.


The event selection for each channel is optimized through a minimization of the expected 
statistical uncertainty on the cross~section using Monte Carlo (MC). 
Signal \ttbar\ events are required to have one isolated electron or muon for the
$\ell \tau$ channel 
or two isolated oppositely charged 
leptons for the $\ell \ell$ channels.
At least one jet is required to have $p_T>30$~GeV. 
All channels, except for $e\mu$, which has the best signal over background ratio,
require another jet with $p_T>20$~GeV.
Jets are accepted in the region $|\eta|<2.5$. 
Leptons are required to have $p_T>15$~GeV in the $\ell \ell$ 
channels. A muon in the $\mu\tau$ channel is required to
have $p_T>20$~GeV and 
an electron in the $e \tau$ channel $p_T>15$~GeV.
Tau leptons are required to have $E_T>10$, $5$, or $10$~GeV for $\tau-$type 1, 2 or 3 respectively.
Muons are accepted in the region $|\eta| < 2.0$, while electrons must be 
within $|\eta|<1.1$ or $1.5<|\eta|<2.5$.
In the $\ell \tau$ channels, events containing any additional 
isolated electron or muon passing the selection criteria used in the $\ell \ell$ channel are 
rejected in order to reduce $Z/\gamma^{\ast} \rightarrow \ell^+ \ell^-$ background and to ensure 
that the $\ell \tau$ channels have no overlap with the $\ell \ell$ channels. 
Furthermore, if more than one $\tau$ lepton is found in an event, only the one
with highest $\tau$ probability (highest {\it NN}$_{\tau}$~\cite{nntau} value) is kept for further analysis.

The selection on $\met$ is crucial for reducing the otherwise large
background from $Z/\gamma^{\ast} \rightarrow \ell^+ \ell^-$. This background is
particularly important in the  $ee$, $\mu\mu$ and $\ell \tau$ channels.
Due to different resolutions in electron energies and muon momenta,
optimization of selections leads to different criteria for the four
channels.
In the $ee$ channel, events with dielectron invariant mass of
$M_{ee}<15$~GeV or $84<M_{ee}<100$~GeV
are rejected. 
For $M_{ee}>100$~GeV ($15<M_{ee}<84$~GeV),
they are required to have $\met > 35$~GeV ($\met > 45$~GeV).
The final selection in the $e\mu$ channel requires 
the scalar sum of the most energetic (leading) lepton $p_T$ and the $p_T$ of the single jet (two 
most energetic jets) to be $H_T>105$~GeV ($H_T>115$~GeV).
This requirement rejects the largest backgrounds in this
final state, which arise from $Z/\gamma^{\ast} \rightarrow \tau^+ \tau^-$
and diboson production. 
In the $\mu\mu$ channel, events are required to have $\met > 40$~GeV.
The dimuon invariant mass $M_{\mu\mu}$ must be larger than 30~GeV.
To reduce $Z/\gamma^{\ast} \rightarrow \mu^+\mu^-$ background, we define
a likelihood ratio variable
based on the per-event \met\ probability distribution, calculated from the expected resolution on \met\
and the energies of electrons, muons and jets.  This \met\ likelihood ratio variable is required to be 
larger than 5.
For  $\ell \tau$ channels, events are required to have
$15< \met<200$~GeV. To reduce the multijet background, a two dimensional selection 
is applied in the ($\Delta \phi(\met,\ell),\met$)
plane where $\Delta \phi(\met,\ell)$ is the difference between the azimuthal angle
of the \met\ direction and of the lepton:
$\Delta \phi(\met,e) > 2.2 - 0.045 \times \met$(GeV)
in the $e \tau$ channel and $\Delta \phi(\met,\mu) > 2.1 - 0.035 \times \met$(GeV) 
in the $\mu \tau$ channel. Furthermore,
in the  $e \tau$ channel, events with electrons and 
$\met$ collinear are rejected by requiring $\cos(\Delta \phi(\met,e)) < 0.9$. 
In the $\mu \tau$ channel, events with a second non-isolated
muon are rejected if the invariant mass of the two muons lies in the mass range
$70 < M_{\mu\mu}< 100 $~GeV. 
The final selection in the $\ell \tau$ channels requires at least one 
$b$-tagged jet.


The acceptance and efficiency for the \ttbar\ signal are derived from a combination of 
MC simulation and data.
Top quark pair production is simulated
using the \alpgen~\cite{alpgen} matrix element generator, assuming \mtop=170~GeV.
These events are processed through \pythia~\cite{pythia}
to simulate fragmentation, hadronization and particle decays
and then passed through a \geant~\cite{geant} based simulation
of the D0 detector. 
Data events from random $p\bar{p}$ crossings are superimposed on MC generated events
to reproduce detector noise and luminosity dependent effects in data.
The same reconstruction process is applied to both data and MC events 
to determine the selection efficiencies.
Lepton trigger and identification efficiencies, as well as lepton momentum 
resolution, are derived from $Z/\gamma^{\ast} \rightarrow \ell^+
\ell^-$ data by strictly identifying one charged lepton as tag and using the
other charged lepton as a probe. The efficiencies are studied in different detector regions
and as a function of the number of jets.
The lepton and jet reconstruction efficiencies, as well as the lepton, jet energy and \met\
resolutions in the MC are adjusted to the values measured in data.

Background contributions are also determined from a combination of MC simulation
and data. The selection efficiencies for the $Z/\gamma^{\ast}$ and $W+$jets backgrounds 
are estimated using MC samples generated by \alpgen\ 
interfaced with \pythia\ while for diboson production they are estimated using
\pythia. The $Z/\gamma^{\ast}$ and diboson
processes are generated at leading order (LO) and are normalized to
the next-to-next-to-leading order (NNLO) inclusive
cross~section and to the next-to-leading order (NLO) 
inclusive cross~sections, respectively~\cite{campbell-ellis,mcfm}.
As the $p_T$ distribution of the $Z$~boson is not well described in the \alpgen\ simulation,
the $p_T$ spectrum was reweighted to reproduce that in $Z \rightarrow e^+e^-$ data 
in the different jet multiplicities.

In the $\ell \tau$ channel, the simulated inclusive background from $W +\ge 2$~jet events is
normalized by fitting the transverse mass distribution~\cite{smith} 
of the isolated lepton and \met\ to data.
We estimate the multijet background from data using events having an electron or muon and a $\tau$ 
lepton of the same-charge (after subtracting contributions from $W$ and same-charge $t \bar t$ MC events).
The $t \bar t$ contributions to the same-charge sample result either from a jet
reconstructed as a $\tau$ lepton or from a misidentification of the charge of the $\tau$ lepton.
Contributions from $Z/\gamma^{\ast}$ and diboson events to the same-charge sample are negligible.

In the $\ell \ell$ channel, the instrumental background is also determined from data. 
False electrons can arise from
jets comprised of an energetic $\pi^0$ or $\eta$, and an overlapping track
from $\gamma \to e^+e^-$ conversion. In the $ee$ and $e\mu$ channels, the
background from false electrons is fitted to the distribution of
the electron likelihood discriminant in the data as done in Ref.~\cite{Abachi:2007:prd}.
The shape of the electron likelihood is determined for true electrons in a $Z/\gamma^{\ast} \rightarrow e^+e^-$
data sample. The shape of the electron likelihood for background electrons is then determined
using a data sample with low $\met$ dominated by false electrons.
An isolated muon can be mimicked by a muon in a jet when the jet is not reconstructed.
We measure the fraction $f_{\mu}$ of muons that appear isolated in a sample enriched 
in semileptonic decays of heavy flavor quarks and in pion or kaon semileptonic in-flight decays.
In this sample, one of the muons is required not to be isolated while the second serves as 
a probe.
In the $\mu\mu$ channel the number of events with a false 
isolated muon that contribute to the final sample is evaluated as in Ref.~\cite{Abachi:2007:prd}.
In the $e\mu$ channel, the contribution from events with a true electron and a false 
isolated muon is given by the number of events in a sample without a muon isolation requirement 
(where the electron and the muon have the same charge) multiplied by the
rate $f_{\mu}$ introduced above.
Although the $Z/\gamma^{\ast} \rightarrow \ell^+ \ell^-$ processes do not lead to high $p_T$ 
neutrinos, they can have large $\met$ from mismeasurements.
The $\met$ spectra from $Z/\gamma^{\ast} \rightarrow \ell^+ \ell^-$
data and the MC agree well, after jet, electron and muon
resolutions are adjusted in the MC to match the resolutions
observed in data.

In the $\ell \tau$ channel, instrumental background can arise from
a candidate electron that does not satisfy electron selection criteria 
but can mimic the signatures of the type~2 $\tau$ lepton.
To discriminate between the $\tau$-type~2 leptons and electrons,
we use another neural network ({\it NN}$_{\rm e}$)~\cite{nntau} along with {\it NN}$_{\tau}$.
The {\it NN}$_{\rm e}$ neural network relies on a subset of the input variables
to {\it NN}$_{\tau}$ and on other variables based on the properties of the electromagnetic clusters
and on the correlation between them and those of the leading track of the $\tau$ lepton. 
In addition, in the $e \tau$ channel, $\tau$ lepton candidates with track $\phi<0.02$~radian 
from the nearest border of the calorimeter module are removed since they are more likely to come 
from misreconstructed electrons. A $\tau$ lepton can also be mimicked by a jet. 
The corresponding rate for such misidentification is determined through
a correction factor from a comparison of $W+$jets MC samples to
$e$+jets data, where the estimated contribution from multijet events as well as 
from $Z \rightarrow e^+e^-$, $Z \rightarrow \tau^+ \tau^-$ and $t \bar t$
have been subtracted via MC. This correction factor is then applied to the $W$+jets and
$t \bar t \rightarrow \ell$+jets samples. 

\begin{figure}[!htb]
\includegraphics[scale=0.2]{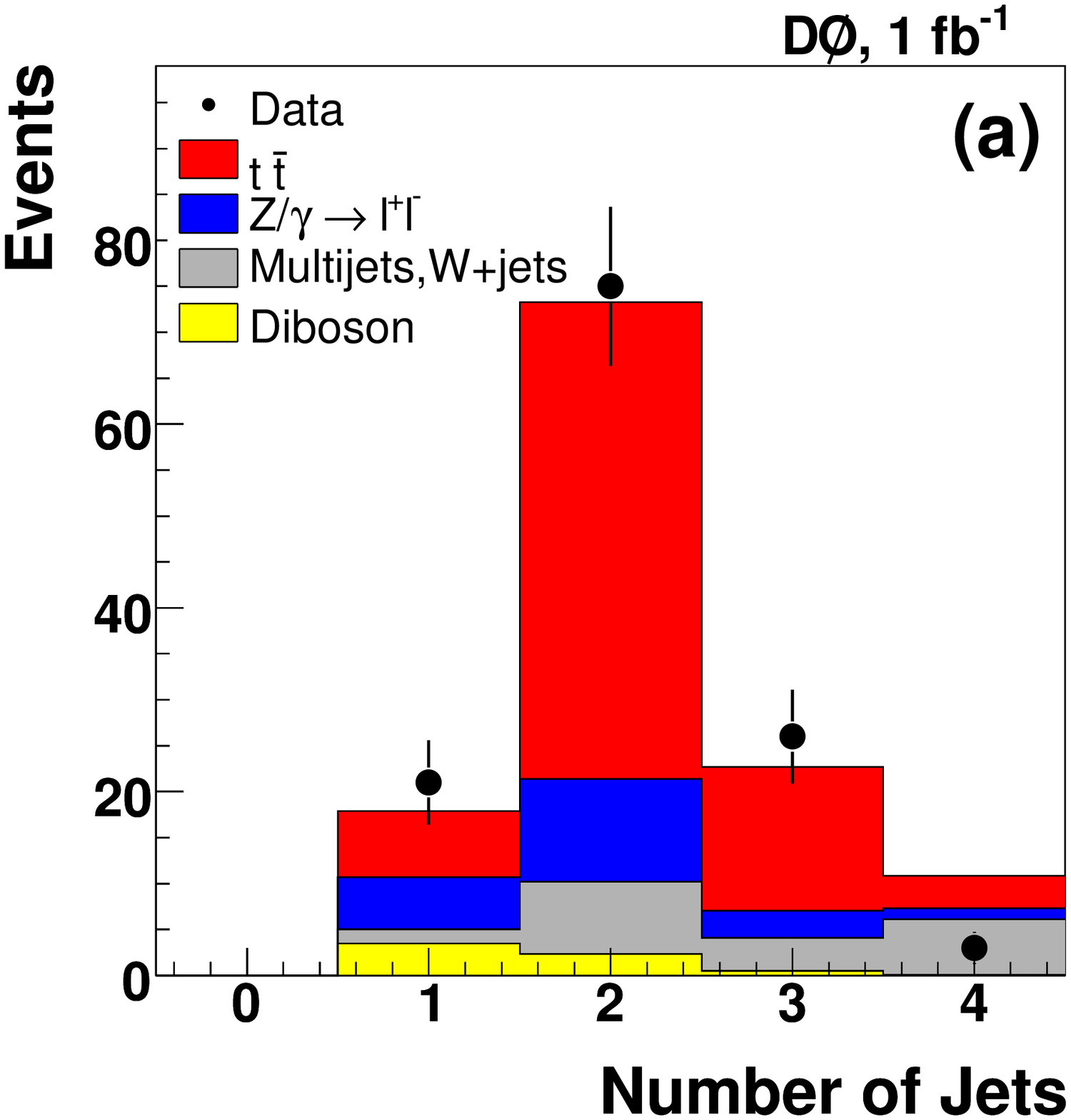}
\includegraphics[scale=0.2]{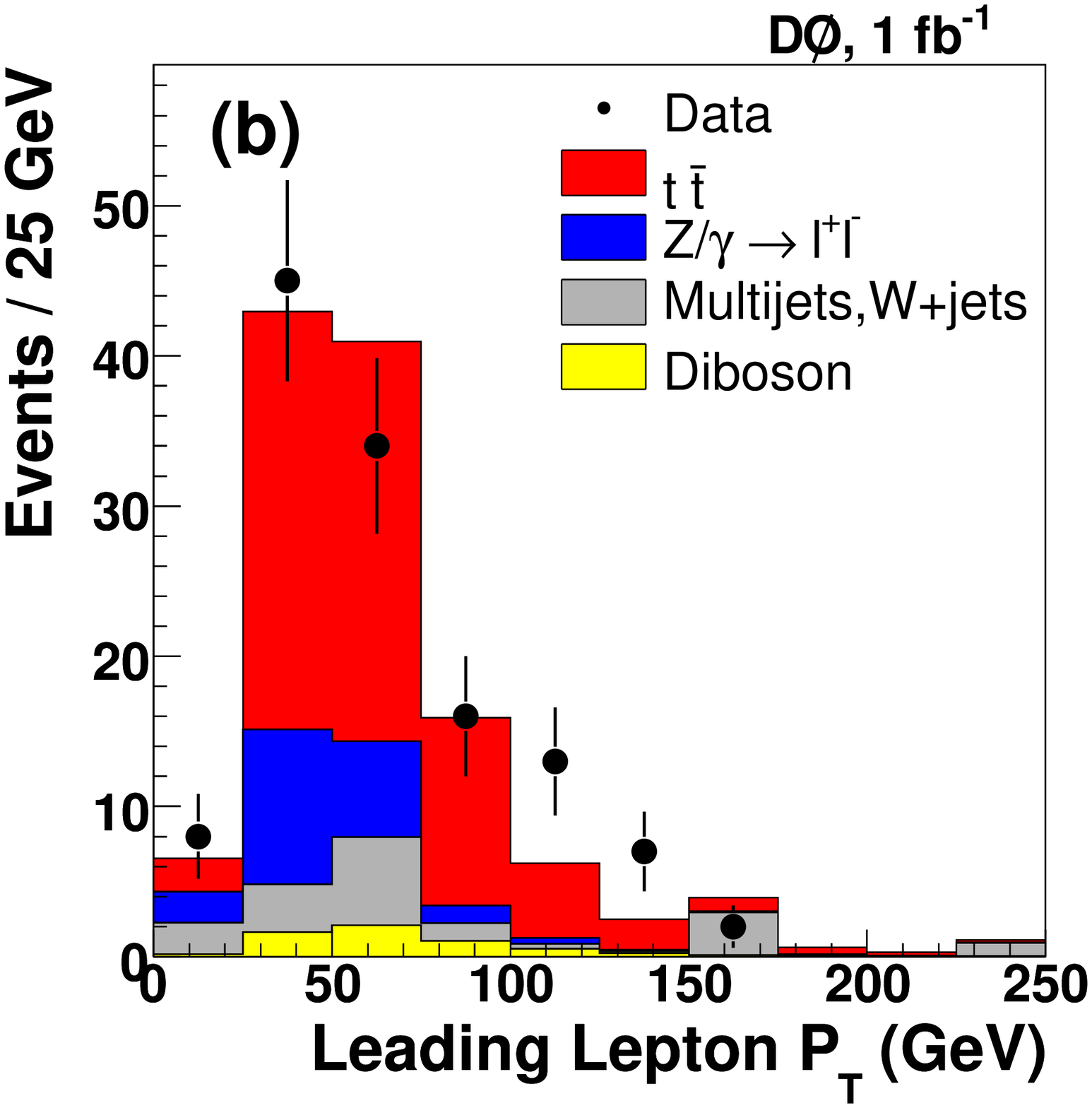}
\includegraphics[scale=0.2]{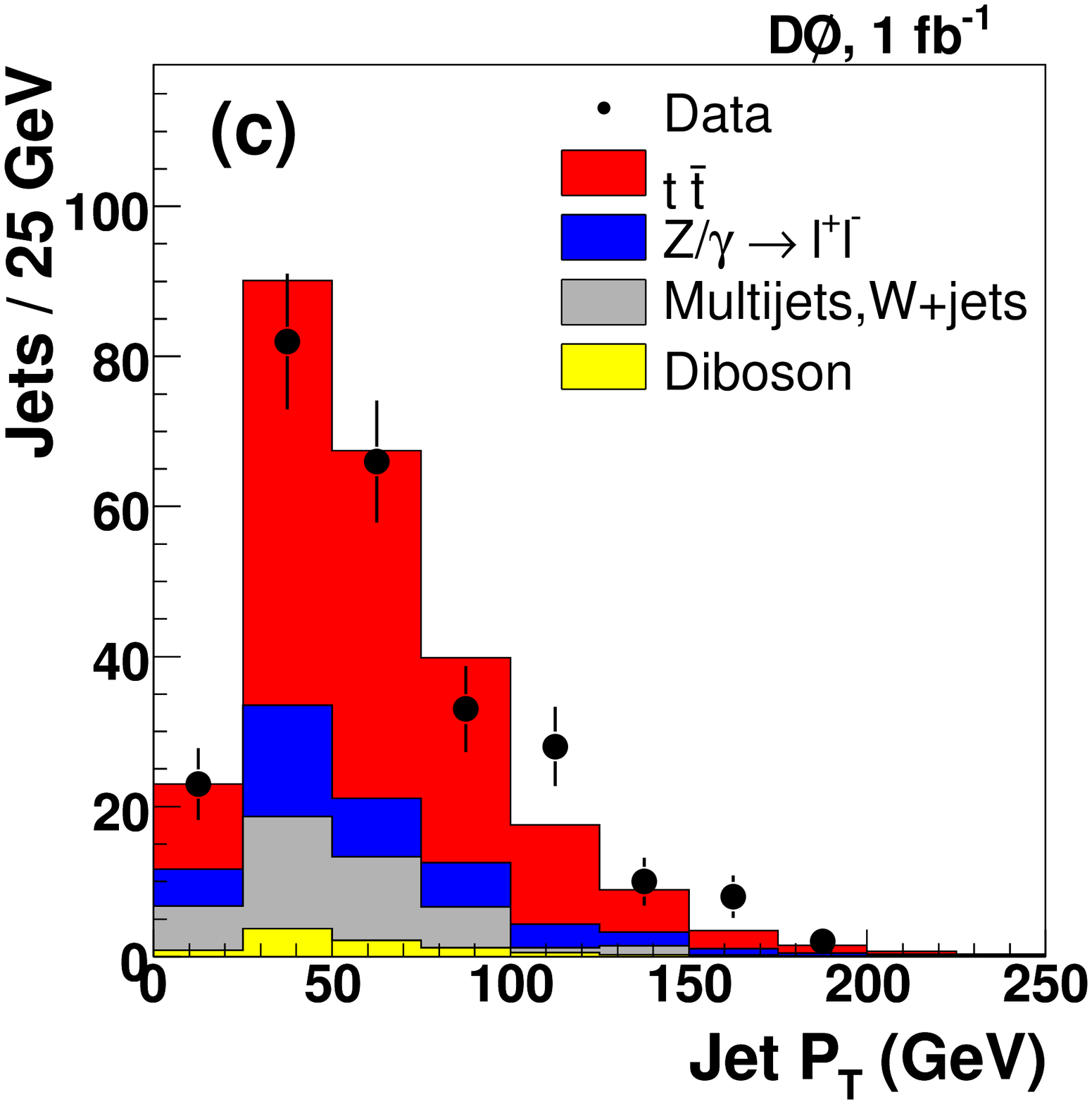}
\includegraphics[scale=0.2]{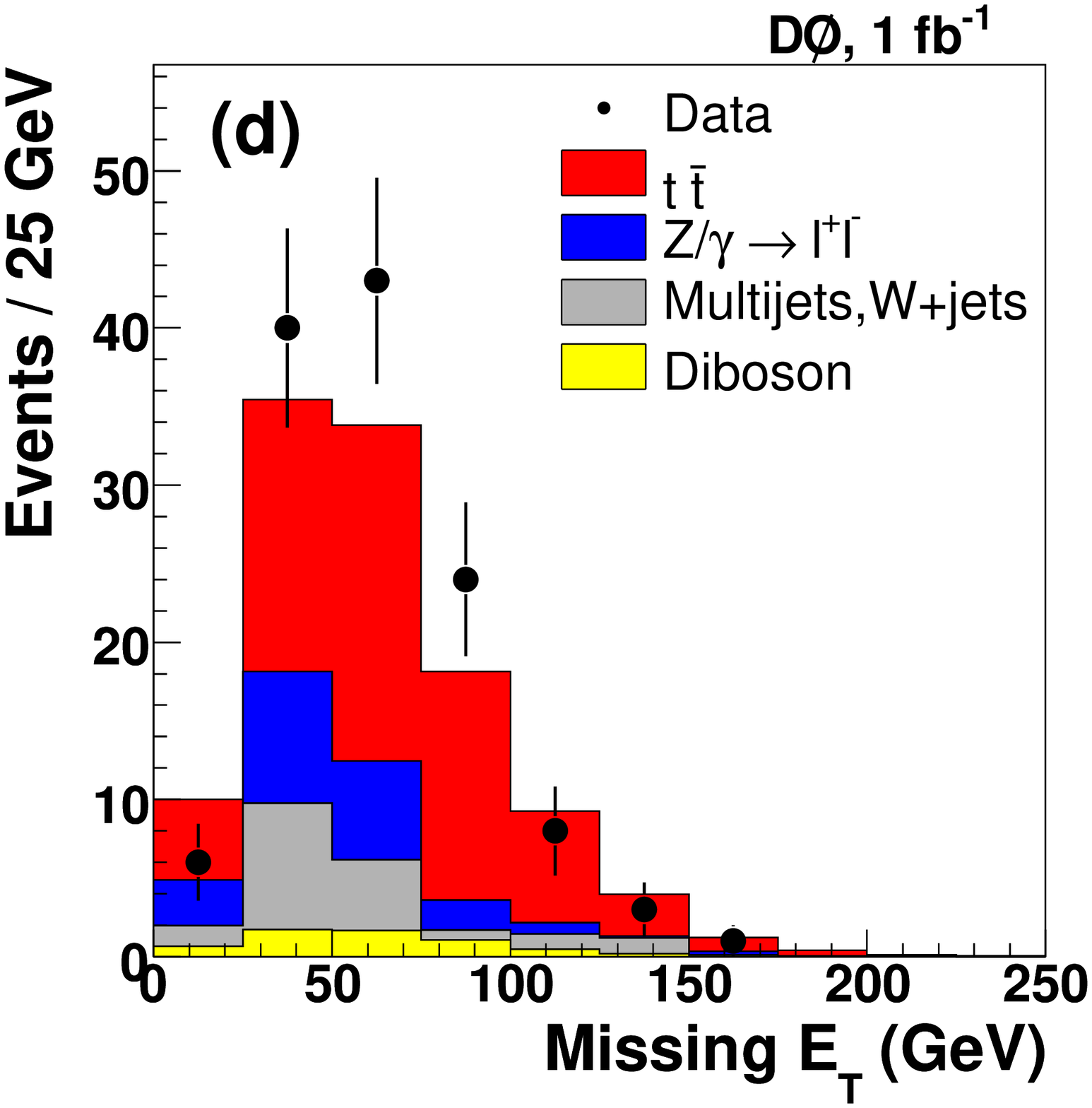}
\caption{Expected and observed distributions for the combined $\ell\ell$ and 
$\ell\tau$ channels for events with $ \ge 1$ jet ($e\mu$) or $ \ge 2$ jets
($ee$, $\mu\mu$, $\ell\tau$) following all selections for (a) the
number of jets per event,
(b) leading lepton \pt,
(c) jet \pt,
and (d) \met.
The \ttbar\ contribution is normalized to the cross section measured in this analysis.}
\label{fig:mcdata}
\end{figure}

\begin{figure}[!htb]
\includegraphics[scale=0.2]{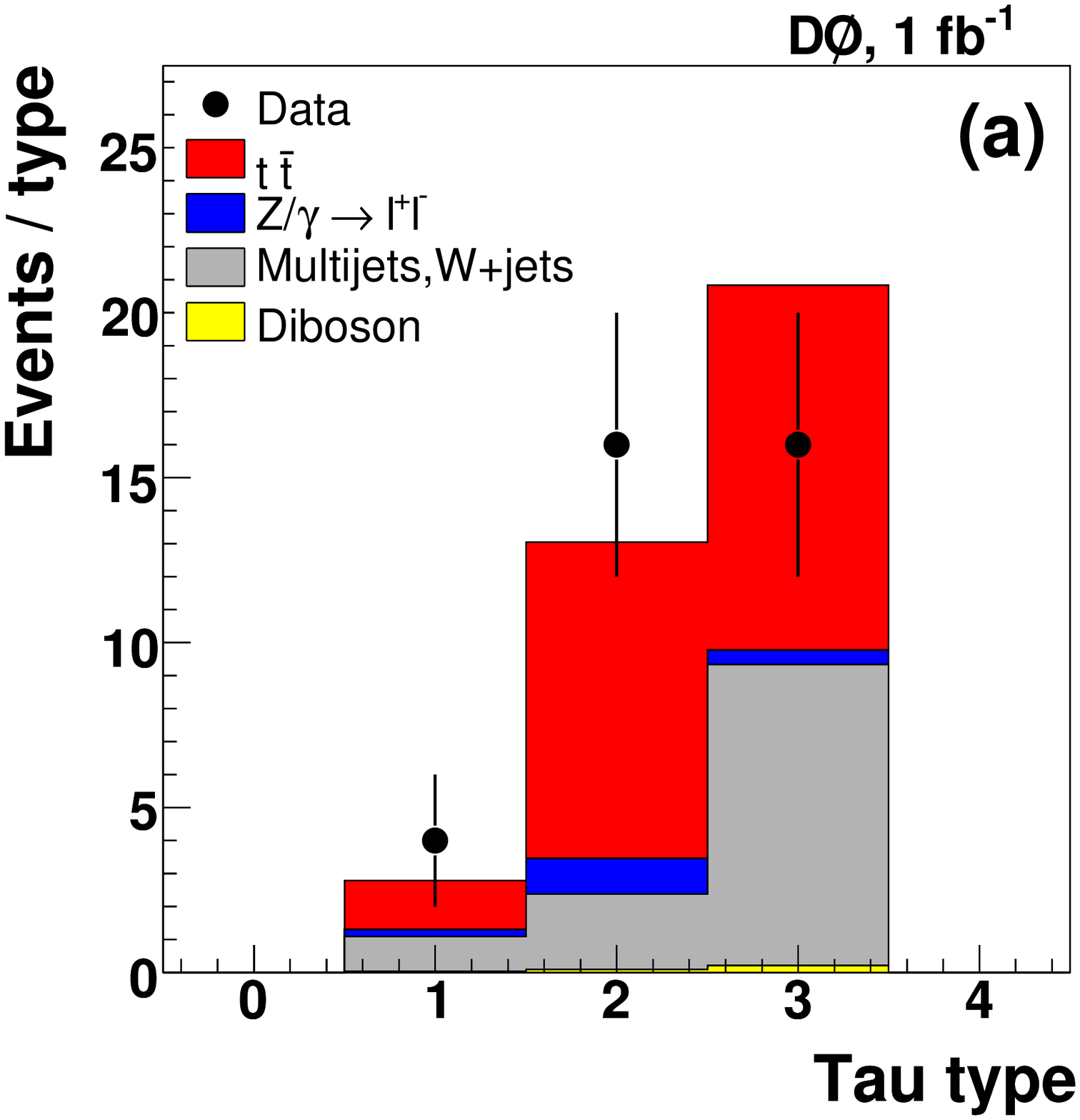}
\includegraphics[scale=0.2]{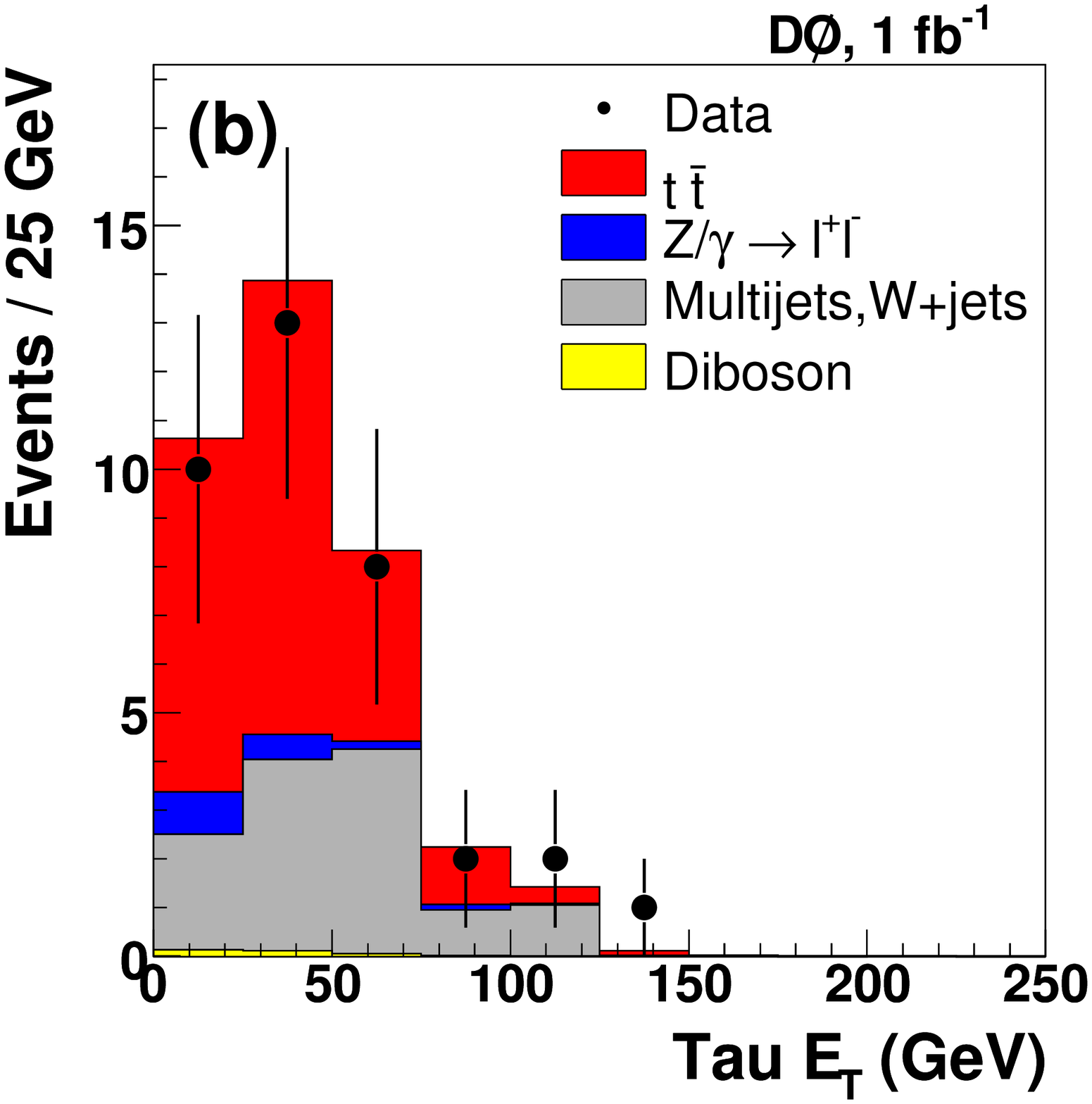}
\caption{Expected and observed distributions in the
$\ell\tau$ channel for (a) the $\tau$-type and (b) $E_T$ of the $\tau$ lepton.
The \ttbar\ contribution is normalized to the cross section measured in the  $\ell\tau$ channel.}
\label{fig:mcdataltau}
\end{figure}

The expected number of background and
signal events and the number observed in data as well as the selection 
efficiencies and luminosities are summarized for all channels in Table~\ref{tab:yields}.
Figure~\ref{fig:mcdata} shows the expected and observed distributions
for several observables in the combined 
$\ell\ell$ and $\ell\tau$ channels.
Figure~\ref{fig:mcdataltau} shows distributions in $\tau$-types
and $E_T$ of the $\tau$ lepton in the $\ell\tau$ channels.
\begin{table*}[ht]
\begin{center}
\caption{Expected number of background and signal events, observed number of events in data, selection efficiencies
and luminosities for all dilepton channels. Uncertainties include both statistical and systematic contributions (excluding luminosity uncertainty of $6.1\%$~\cite{lumi}).
The signal efficiency is quoted for \mtop=170~GeV and the expected number of signal events for $\sigma_{t\bar{t}}=7.9$~pb \cite{kidonakis}.}
\begin{tabular}{l||l|l|l|l|l|l|}\hline \hline
Channel & $ee$ & $e\mu$ (1 jet) & $e\mu$ ($\geq$ 2 jets) & $\mu\mu$ & $e\tau$ & $\mu\tau$ \\[2pt] \hline
Luminosity ($\rm pb^{-1}$) & 1074 
			& 1070 & 1070 
			& 1009 
			& 1038 & 996 \\[2pt] \hline \\[-10pt]
$Z/\gamma^{\ast}$       & 2.4$^{+0.6}_{-0.5}$ 
			& 5.5$^{+0.7}_{-0.8}$ & 5.4$^{+0.9}_{-1.0}$
			& 5.6$^{+1.0}_{-1.2}$ 
			& 0.6$^{+0.1} _{-0.1}$ & 1.2$^{+0.3} _{-0.2}$ \\[2pt]
$WW/WZ/ZZ$              & 0.5$^{+0.1}_{-0.1}$ 
			& 3.1$^{+0.7}_{-0.7}$ & 1.4$^{+0.4}_{-0.4}$ 
			& 0.6$^{+0.1}_{-0.1}$ 
			& 0.2$^{+0.0}_{-0.0}$ & 0.2$^{+0.0}_{-0.0}$ \\[2pt]
Multijet/$W$+jets       & 0.6$^{+0.4}_{-0.4}$ 
			& 0.9$^{+0.3}_{-0.2}$ & 2.6$^{+0.6}_{-0.5}$ 
			& 0.2$^{+0.2}_{-0.2}$ 
			& 3.6$^{+1.8} _{-1.8}$ & 8.8$^{+2.8}_{-2.8}$ \\[2pt] \hline \\[-10pt]
Total background        & 3.4$^{+0.7}_{-0.6}$ 
			& 9.5$^{+1.0}_{-1.1}$ & 9.4$^{+1.2}_{-1.2}$ 
			& 6.4$^{+1.9}_{-1.1}$ 
			& 4.4$^{+1.8}_{-1.8}$ & 10.2$^{+2.9}_{-2.9}$\\[2pt] \hline \\[-10pt]
Signal efficiency (\%)  & 1.3$^{+0.1}_{-0.1}$ 
			& 1.0$^{+0.0}_{-0.0}$ & 3.9$^{+0.0}_{-0.0}$ 
			& 1.1$^{+0.0}_{-0.0}$ 
                        & 0.23$^{+0.1}_{-0.1}$ & 0.28$^{+0.1}_{-0.1}$\\[2pt]
Expected signal         &  11.2$^{+0.8}_{-0.8}$ 
			& 8.6$^{+1.1}_{-1.1}$ & 35.2$^{+2.6}_{-2.7}$ 
			& 8.8$^{+0.8}_{-0.8}$ 
			& 10.3$^{+1.1}_{-1.1}$ & 12.2$^{+1.1}_{-1.1}$\\[2pt] \hline \\[-10pt]
Total expected	        & 14.6$^{+1.0}_{-1.0}$ 
			& 18.0$^{+1.4}_{-1.6}$ & 44.6$^{+3.4}_{-3.6}$   
			& 15.1$^{+1.5}_{-1.6}$ 
			& 14.7$^{+2.0}_{-2.0}$ & 22.3$^{+3.1}_{-3.1}$ \\[2pt] \hline \hline
Data	                & 17 
			& 21 & 39 
			& 12 
			& 16 & 20 \\ \hline \hline
\hline
\end{tabular}
\label{tab:yields}
\end{center}
\end{table*}


The systematic uncertainty on the measured \ttbar\ production cross section
in the dilepton channel is
obtained by varying the efficiencies and background contributions within their
uncertainties, taking all correlations among the different channels
and background contributions into account.
The statistical uncertainties on MC and backgrounds
are treated as uncorrelated among channels, while other sources of
systematic uncertainty are treated as correlated.
The dominant systematic uncertainties
are summarized in Table~\ref{tab:syst} for individual channels and in Table~\ref{tab:syst_comb_nuis}
for the combination of channels.

The systematic uncertainties on trigger efficiencies \mbox{($\sim \!2\%$ of the cross section)} 
are derived from data. Various sources of bias are investigated, and the resulting changes in 
trigger efficiencies are included as systematic uncertainties.

The systematic uncertainty for identifying $\tau$ lepton ($\sim \!5.5\%$ of the $\ell \tau$ cross section) 
arises dominantly from the uncertainty on the data to MC 
agreement and from the statistical uncertainty on the correction factor for jets mimicking $\tau$ leptons.
The systematic uncertainty for the $\tau$ lepton energy scale ($\sim \!6\%$ of the $\ell \tau$ cross section) 
is estimated from the calorimeter's response to single pions~\cite{nntau}.

The systematic uncertainties from the reconstruction
and resolution of jets ($\sim \!1\%$) are determined from the uncertainty on the data/MC correction factors.
The uncertainty on the calibration of jet energy ($\sim \!4\%$) is
propagated to the predicted background and to the efficiency for
$t \bar t$ signal.

The uncertainty on $b$ tagging specific to the  $\ell \tau$ channel ($\sim \!4.5\%$)
is evaluated by shifting the jet tagging probability within its uncertainty.
The flavor dependent uncertainties are evaluated by
changing the parametrization of the tagging probability for different types
of jets ($b$, $c$ and light jets).

The uncertainty on theoretical modeling of $t\bar t$ production ($\sim \!5\%$) is
estimated by comparing the acceptance of the two
MC programs, \pythia\ and \alpgen.
The full difference in the final result is quoted as the systematic uncertainty.
Half of the difference between unity and the ratio of the NLO diboson cross section
to the LO diboson cross section (used to scale the diboson cross sections in \pythia)
is taken as a systematic uncertainty for the diboson background. The systematic uncertainty
on the normalization of the $Z/\gamma^{\ast}$ background
is estimated by propagating the uncertainty on the $p_T$ reweighting function of the $Z$ boson.

The systematic uncertainties on electron background
in the $e\mu$ and $ee$ channels are evaluated using the shape dependence of the
electron likelihood discriminant on electron $p_T$ and the detector
occupancy (number of jets).

Other smaller sources of systematic uncertainties \mbox{($\sim \!2.5\%$)} arise from
vertex identification, parton distribution functions and $\met$ modeling.
The luminosity uncertainty \mbox{($\sim \!6\%$)}~\cite{lumi} on the cross sections
is evaluated taking into account both the uncertainty on the predicted number of signal and
background events.

\begin{table*}[ht]
\begin{center}
\caption{Summary of the effects of individual systematic uncertainties on the measured cross section (in pb).}
\begin{tabular}{l|c|c|c|c|c|c}\hline \hline 
Source & $ee$ & $e\mu$ (1 jet) & $e\mu$ ($\geq$ 2 jets) & $\mu\mu$ & $e \tau$ & $\mu \tau$  \\ \hline
Trigger               & +0.0   -0.0 & +0.6    +0.0 & +0.2  -0.0  & +0.6 -0.4  & +0.2    -0.1  &  +0.3    -0.2 \\
Lepton identification & +0.4   -0.4 & +0.5    -0.5 & +0.2  -0.2  & +0.5 -0.4  & +0.3    -0.2  &  +0.2    -0.2  \\
Tau identification    & n/a         & n/a          & n/a         & n/a        & +0.6    -0.5  &  +0.4    -0.3  \\
Tau energy scale      & n/a         & n/a          & n/a         & n/a        & +0.5    -0.4  &  +0.4    -0.4   \\
Jet identification    & +0.1   -0.2 & +0.4    -0.4 & +0.1  -0.1  & +0.2 -0.4  & +0.1    -0.1  &  +0.1    -0.1  \\
Jet energy scale      & +0.6   -0.5 & +0.8    -0.7 & +0.5  -0.5  & +0.6 -0.2  & +0.2    -0.2  &  +0.3    -0.2  \\
$b$ jet identification  & n/a         & n/a          & n/a         & n/a        & +0.4    -0.4  &  +0.3    -0.3  \\
Signal modeling       & +0.4   -0.4 & +1.1    -1.0 & +0.3  -0.3  & +0.3 -0.3  & +1.0    -0.8  &  +0.6    -0.5  \\
Background estimation & +0.2   -0.1 & +0.6    -0.5 & +0.1  -0.1  & +0.4 -0.3  & +0.1    -0.1  &  +0.1    -0.1  \\
False lepton background & +0.3   -0.3 & +0.2    -0.3 & +0.1  -0.1  & +0.1 -0.1  & +1.3    -1.3  &  +1.8    -1.8  \\
Other                 & +0.4   -0.4 & +0.7    -0.7 & +0.2  -0.2  & +0.7 -0.7  & +0.3    -0.3  &  +0.2    -0.2  \\
Total                 & +1.0   -0.9 & +1.9    -1.6 & +0.8  -0.7  & +1.3 -1.1  & +1.9    -1.8  &  +2.1    -2.0 \\ \hline
Luminosity            & +0.8   -0.7 & +1.2    -1.1 & +0.5  -0.5  & +0.7 -0.6  & +0.6    -0.5  &  +0.5    -0.4  \\ \hline \end{tabular}
 \label{tab:syst}
 \end{center}
\end{table*}
\begin{table*}[ht]
\begin{center}
\caption{Summary of the effects of individual systematic uncertainties on the combined cross section (in pb).}
\begin{tabular}{l|c|c}\hline \hline
Source & dilepton ($\ell \ell$) & combined ($\ell \ell$ + $\ell \tau$) \\ \hline
Trigger               & +0.2   -0.1 &  +0.2 -0.1 \\
Lepton identification & +0.2   -0.2 &  +0.2 -0.2 \\
Tau identification    & n/a         &  +0.1 -0.1 \\
Tau energy scale      & n/a         &  +0.1 -0.1 \\
Jet identification    & +0.0   -0.1 &  +0.0 -0.0 \\
Jet energy scale      & +0.4   -0.4 &  +0.3 -0.3 \\
$b$ jet identification  & n/a         &  +0.1 -0.1 \\
Signal modeling       & +0.3   -0.3 &  +0.4 -0.4 \\
Background estimation & +0.2   -0.1 &  +0.1 -0.1 \\
False lepton background & +0.1   -0.1 &  +0.3 -0.3 \\
Other                 & +0.3   -0.3 &  +0.2 -0.2 \\
Total                 & +0.7   -0.6 &  +0.7 -0.6 \\ \hline
Luminosity            & +0.7   -0.5 &  +0.6 -0.5 \\
\hline\end{tabular}
 \label{tab:syst_comb_nuis}
 \end{center}
\end{table*}


Cross sections for individual channels are extracted using a likelihood technique
described in Ref.~\cite{Abachi:2007:prd}. The results are presented in
Table~\ref{tab:result}. All cross sections agree within their uncertainties.
\begin{table}[ht]
\begin{center}
\caption{The measured \ttbar\ cross section at $\sqrt{s}=1.96$~TeV for \mtop=170~GeV.}
\begin{tabular}{c|c}\hline \hline
Channel & $\sigma_{t\bar{t}}$ (pb) \\[2pt] \hline \\[-8pt]
ee         & \fullee \\[2pt]
e$\mu$ ($\geq$ 1 jet)    & \fullemu \\[2pt]
$\mu \mu$  & \fullmumu \\[2pt]
e$\tau$    & 8.9$^{+3.3}_{-2.8}$ (stat) $^{+1.9}_{-1.8}$ (syst) $^{+0.6}_{-0.5}$ (lumi) \\[2pt]
$\mu \tau$ & 6.4$^{+3.1}_{-2.7}$ (stat) $^{+2.1}_{-2.0}$ (syst) $^{+0.5}_{-0.4}$ (lumi) \\[2pt] \hline \hline
\end{tabular}
\label{tab:result}
\end{center}
\end{table}
The combined result is obtained by minimizing the sum of
negative log-likelihood functions from the five channels. All systematic 
uncertainties are incorporated in the fit as ``nuisance
parameters"~\cite{nuisance} that can affect the central value of the cross section.
The result from combining the $ee$, $e\mu$ and $\mu\mu$ ($\ell \ell$) channels is: 
\[
\sigma_{t\bar{t}}=\fullresultll \, \rm pb
\]
and for the $\ell \ell$ and $\ell \tau$ channels combined:
\[
\sigma_{t\bar{t}}=\fullresult \, \rm pb. \nonumber
\]
Both results are derived for \mtop=170~GeV.
These represent the most precise \ttbar\ cross section measurements
published so far in the dilepton channel. 

To improve the statistical uncertainty in the $\ell \tau$ channels, 
the signal acceptance for all the $\ell \tau$ results quoted above includes contributions from \ttbar\ events 
in which the $\tau$ selection is satisfied by jets mimicked $\tau$ leptons.
If we now use only \ttbar\ events that decay specifically to $\ell \tau$ final states, we measure:
\[
\sigma_{t\bar{t}}=7.6^{+4.9}_{ -4.3}\text{(stat)}
		^{+3.5} _{-3.4} (\mathrm{syst})
		^{+1.4} _{-0.9} (\mathrm{lumi}) \mathrm{~pb}.
\]
A measurement of the cross section multiplied by the branching ratio ($\sigma_{t\bar{t}} \times B$) 
has also been performed in the $\ell \tau$ channel using the acceptance from \ttbar\ events
that decay specifically to $\ell \tau$ final states (where only \ttbar\ events which contain a hadronically decaying $\tau$ lepton at generator level are considered).  The expected
contribution from other \ttbar\ events is normalized using the theoretical cross section \cite{kidonakis}.
In the combined $e\tau$ and $\mu\tau$ channels we obtain the value for $\sigma_{t\bar{t}} \times B(\ttbar \to \ell \tau b\bar{b})$:
\[
   \sigma_{t\bar{t}} \times B = 0.13
   ^{+0.09}_{-0.08}\text{(stat)}
   ^{+0.06} _{-0.06} (\mathrm{syst})
   ^{+0.02} _{-0.02} (\mathrm{lumi}) \mathrm{~pb},
\]
for \mtop=170~GeV, which is in good agreement with the SM 
expectation of $0.14 \pm 0.02$ pb~\cite{cacciari,Yao:2006px}.
Dividing the $\sigma_{t\bar{t}} \times B(\ttbar \to \ell \tau b\bar{b})$ measurement by the SM 
expectation, we can set an upper limit on the ratio of 2.3 at 95\% confidence level (CL).

%

The value of quark masses depends on the perturbative QCD renormalization scheme,
and can differ considerably for, e.g., 
pole mass or $\overline{\rm MS}$ mass
definitions~\cite{massdefs}. 
It is therefore important to extract the mass of the top quark through 
a well-defined renormalization scheme. Direct top quark mass measurements
compare measured distributions to distributions simulated by LO MC generators.
Like any LO calculation, these MC generators are not precise enough to fix the 
renormalization scheme, which leads to uncertainty in the input mass definition.
In the present analysis, we extract the mass of the top quark using the measured 
top pair production cross section. This has the advantage
of not relying on simulation of the \ttbar\ signal, except for determining
detection efficiency. The sensitivity to any differences
between the pole mass and the mass used in the MC simulation is
thereby reduced relative to a direct 
mass measurement. We compare our result to fully 
inclusive \ttbar\ cross sections calculated in higher-order QCD that includes soft
gluon resummations, which are currently the most 
complete calculations available. The cross sections are computed using the pole
mass definition for the top quark which is thus 
the parameter extracted here.

We extract the \ttbar\ cross section $\sigma_{t\bar{t}}$ combining the $\ell \ell$ and $\ell \tau$ channels
using the selections described above and different values of the top quark
mass for calculating detection efficiencies in fully simulated 
\ttbar\ events. The result is extracted using the same function as given in Ref.~\cite{cacciari},
$$\sigma_{t\bar{t}} = \frac{1}{m^4_t} [ a + b (m_t-170) + c  (m_t-170)^2 + d  (m_t-170)^3 ]$$
with $a = 6.28727 \times 10^{9}, b = 9.12630 \times 10^{7}, c = 8.38430 \times 10^{5}$ and $d = - 3.898 \times 10^{5}$ and where $\sigma_{t\bar{t}}$ and \mtop\ are in pb and GeV respectively.
   
\begin{figure}[h]
\includegraphics[scale=0.35]{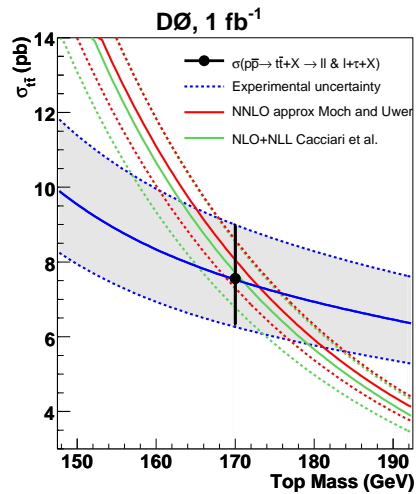}
\caption{Dependence of the experimental and theoretical~\cite{cacciari,moch} \ttbar\ cross section on \mtop. 
The point shows the combination of the $\ell \ell$ and $\ell \tau$ measurements presented in this Letter.}
\label{fig:xs_vs_mass}
\end{figure}

Figure~\ref{fig:xs_vs_mass} compares this
parameterization of the combined measurement with a prediction including
soft gluon resummation effects~\cite{cacciari} and an approximate NNLO computation~\cite{moch}.
For the theoretical computation we plot a 68\%~CL interval that we
determine based on Ref.~\cite{cacciari} or \cite{moch}. The uncertainty from
the ambiguity in the scale of QCD (which are varied from $m_t/2$ to $2 m_t$)
is represented by a likelihood function that is constant
within the ranges given in Ref.~\cite{cacciari} or \cite{moch} and vanishes elsewhere.
The uncertainty due to the parton distribution functions is represented by a Gaussian
likelihood, with rms equal to the uncertainty determined
in Ref.~\cite{cacciari} or \cite{moch}. 
For every value of the mass of the top quark, we form a joint normalized likelihood 
function based on the theoretical likelihoods and on a likelihood
for the measurement constructed from a Gaussian with rms equal to the total 
experimental uncertainty~\cite{prlljets}.
We find $m_t = \mtc \ermtc$~GeV at 68\%~ CL
using Ref.~\cite{cacciari} and 
$m_t = \mtm \ermtm$~GeV at 68\%~CL using Ref.~\cite{moch}. 
These values are in agreement with the
current world average of $m_t = 172.4 \pm 1.2\ {\rm
  GeV}$~\cite{top_worldaverage}, indicating that any deviation of the
  directly measured mass from the true pole mass of the mass of the top quark is $ \lsim 10\ {\rm GeV}$ at 
68\%~CL.

In summary, we described in this Letter the measurement of the \ttbar\ cross section 
in the dilepton and lepton+$\tau$ channels using approximately 1~fb$^{-1}$ of D0 data. 
The combined cross section is measured to be: 
$\sigma_{t\bar{t}}=\fullresult$~pb for a mass of the top quark of \mtop=170~GeV, in agreement 
with the QCD prediction. We measured 
$\sigma_{t\bar{t}} \times B(\ttbar \to \ell \tau b\bar{b})= 0.13
   ^{+0.09}_{-0.08}\text{(stat)}
   ^{+0.06} _{-0.06} (\mathrm{syst})
   ^{+0.02} _{-0.02} (\mathrm{lumi}) \mathrm{~~pb}$ which agrees with the 
SM expectation. Using both the $\ttbar$ cross section measurement and the theoretical prediction,
we extract the mass of the top quark: $m_t=\mtm \ermtm$~GeV which is consistent with the mass of the top quark 
from direct measurements.

%
We thank the staffs at Fermilab and collaborating institutions, 
and acknowledge support from the 
DOE and NSF (USA);
CEA and CNRS/IN2P3 (France);
FASI, Rosatom and RFBR (Russia);
CNPq, FAPERJ, FAPESP and FUNDUNESP (Brazil);
DAE and DST (India);
Colciencias (Colombia);
CONACyT (Mexico);
KRF and KOSEF (Korea);
CONICET and UBACyT (Argentina);
FOM (The Netherlands);
STFC (United Kingdom);
MSMT and GACR (Czech Republic);
CRC Program, CFI, NSERC and WestGrid Project (Canada);
BMBF and DFG (Germany);
SFI (Ireland);
The Swedish Research Council (Sweden);
CAS and CNSF (China);
and the
Alexander von Humboldt Foundation (Germany).
%



\begin{thebibliography}{99}
\bibitem{pseudorap}
  The pseudorapidity is defined as $\eta = - \ln [\tan(\theta/2)]$ where $\theta$ is the polar
  angle with respect to the proton beam.
\end{thebibliography}
\end{document}